\documentclass[prb,twocolumn,showpacs,preprintnumbers,amsmath,amssymb,superscriptaddress]{revtex4-1}

\usepackage{dcolumn}
\usepackage{bm}
\usepackage{amsmath,graphicx}
\usepackage{color}

\usepackage{bm}
\usepackage{url}

\begin{document}
\title{Universal fluctuation of polygonal crack geometry in solidified lava}

\author{Yuri Akiba}
\affiliation{Department of Environmental Sciences, University of Yamanashi, 4-4-37, Takeda, Kofu, Yamanashi 400-8510, Japan}

\author{Aika Takashima}
\affiliation{Department of Environmental Sciences, University of Yamanashi, 4-4-37, Takeda, Kofu, Yamanashi 400-8510, Japan}

\author{Hiroyuki Shima}
\email{hshima@yamanashi.ac.jp}
\thanks{(Correspondence author)} 
\affiliation{Department of Environmental Sciences, University of Yamanashi, 4-4-37, Takeda, Kofu, Yamanashi 400-8510, Japan}

\date{\today}

%***********
\begin{abstract}
Outcrops of columnar joints made of solidified lava flows are often covered by semi-ordered polygonal cracks. The polygon diameters are fairly uniform at each outcrop, but their shapes largely vary in the number of sides and internal angles. Herein, we unveil that the statistical variation in the polygon shape follows an extreme value distribution class: the Gumbel distribution. The Gumbel law was found to hold for different columnar joints, regardless of the locality, lithologic composition, and typical diameter. A common distribution for columnar joints implies a new universal class that may integrate the polygonal crack networks observed on the surface of various fractured brittle materials.
\end{abstract}
%***********

\maketitle

% %**********************************************************
\section{Introduction}
% %**********************************************************

\begin{figure}[bbb]
\centering
\includegraphics[width=0.47\textwidth]{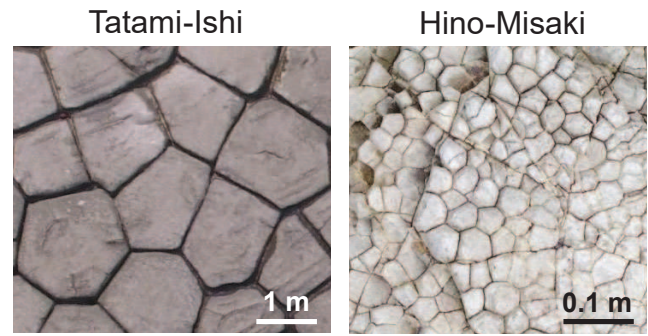}
\caption{Aerial photos of the polygonal crack network observed at the outcrop of columnar joints in Japan:
(a) Tatami-Ishi in Okinawa (O-site) made from andesite and (b) Hino-Misaki in Shimane (Sm-site) made from rhyolite.}
\label{fig_01}
\end{figure}

Nature often fascinates our eyes with its sculptural beauty.
A familiar example is the autonomous formation of polygonal patterns, which is observed in a variety of fields
such as wasp nests in biology \cite{Nazzi2016,DJeong2019},
bubble clusters in fluids \cite{Weaire1984,ZHuang2016}, and
terrestrial fragmentation in geology \cite{Kessler2003,Sletten2003,Domokos2020}.
Under laboratory conditions, polygonal crack patterns occur in various materials including 
calcium carbonate mixture \cite{Nakahara2006,Akiba2019}, alumina-water mixture \cite{Shorlin2000},
metal thin films \cite{Yu2019}, and frozen impacted water droplets \cite{Ghabache2016}.
A spectacular and eye-catching example of polygonal pattern formation in geology is columnar jointing \cite{DeGraff1987,Budkewitsch1994}.
It is an ordered assembly of long slender prismatic cracks that spontaneously develop in volcanic rock.
Self-organization of the spatially periodic prismatic cracks is the result of the thermal contraction 
of solidified lava flow followed by its shrinkage fracture that penetrates inward gradually \cite{Hardee1980}.
In most fracturing phenomena, the direction of cracks is greatly influenced
by the spatial variation of the strength distribution of the material \cite{Wittel2004,Halasz2017}, 
resulting in a random crack network often found in aged concrete walls and windowpanes.
Compared to those random fracturing, the ordered cracks observed on columnar joints seem to be extremely curious, 
and thus have long arouse interest in elucidating the driving force that autonomously creates a collection of elongated columns 
with polygonal cross sections of quasi-equal diameter (see Fig.~\ref{fig_01}).

%----------------
%----------------

It is broadly accepted that the mechanism of columnar joint formation is attributed to 
the maturation process of shrinkage cracks penetrating inward,
as theoretically derived from the principle of the maximum energy release rate \cite{Jagla2002,Jagla2004}.
The formation starts from a superficial random crack network that has occurred during the initial cooling stage 
on the solidifying lava surface.
In the random network, intercrack spacings are rather inhomogeneous, whereby 
most intercrack junctions are T-shaped \cite{Aydin1988}.
As the lower parts of the lava cool, the cracks penetrate the solidified lava body while intercrack junctions transform gradually into Y-shaped; as a consequence, 
the shape of column sections approaches quasi-equal-sized polygons with a preference for hexagons \cite{Hofmann2015}.
Simultaneously, adjacent crack fronts interact with each other;
{\it i.e.,} when one crack grows, the contraction stress around it is relaxed, 
inhibiting the growth of adjacent cracks. 
As a result, a few adjacent columns merge into one, and the cross-sectional area increases 
intermittently with crack growth \cite{Bahr2009}.
A sequence of such transformation and coarsening processes leads to the maturation of the polygonal pattern 
toward a quasi-uniform one with quasi-equal crack spacings.

It should be noted that the polygonal crack network in real columnar joints is not ideally regular 
but exhibits a mixture of different edge lengths and internal angles, involving non-hexagons 
in a considerable proportion \cite{AkibaESS2021}.
This irregularity, which is partly a remnant of the initial random cracking at the early stage,
causes the geometric deviation of cross-sectional polygons from regular counterparts.
Nevertheless, maturation is thought to suppress the degree of deviation 
so that the diameter of each column is comparable to or not much smaller than 
the characteristic crack spacing determined by the mechanical energy balance \cite{Grossenbacher1995,Hofmann2011}
as well as lava composition \cite{Lescinsky2000}.
The two competing effects, {\it i.e.,} the persistence of irregularity and the maturing evolution toward equidiameter columns,
raise the question as to what kinds of statistical properties dominate the polygon geometry 
over the real columnar joints.

In this article, we demonstrate that the statistical fluctuation in the polygon geometry
observed at real columnar joints falls into a specific class of probability distribution.
Surprisingly, this fact is true for all the data obtained from different investigation sites, despite large differences in
locality, lithologic composition, and typical column size.
The robustness of the probability distribution implies
the existence of a previously unknown universal class that governs the geometric fluctuation in polygonal crack patterns.
%\color{red}
A physical interpretation of these results can be obtained from numerical simulations based on two-dimensional Voronoi tessellation.
%\color{black}
We also discuss the possible relevance of our findings to non-Gaussian fluctuation phenomena,
characterized by the generalized Gumbel law, which are ubiquitous in various complex systems.

% %**********************************************************
\section{Methods}
% %**********************************************************
\subsection{Field observation of columnar joint}
% %**********************************************************

\begin{figure}[ttt]
\centering
\includegraphics[width=0.3\textwidth]{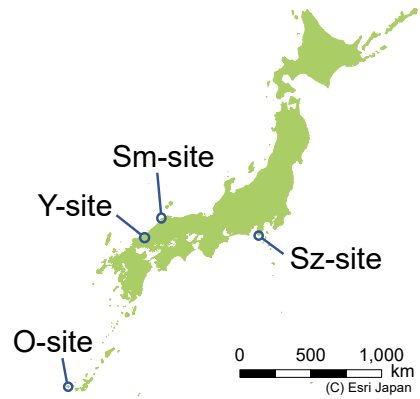}
\caption{Location map for the four investigation sites: 
Okinawa (O-site), Yamaguchi (Y-site), Shimane (Sm-site), and Shizuoka (Sz-site) prefectures in Japan.}
\label{fig_02}
\end{figure}

Our field observation of columnar jointing was conducted in Japan, which is one of the most volcanically active countries worldwide. In fact, various columnar joints with diversity 
in both lithological character and column diameter are distributed over more than 60 locations. 
We investigated four specific sites with different lithologies, as depicted in Fig.~\ref{fig_02};
the corresponding polygonal fractures on the exposed surfaces were photographed from above by a drone.
Figure \ref{fig_01} shows two exemplary aerial photos taken at O-site and Sm-site.
From the photos, it follows that the entire outcrop is occupied by polygonal cracks,
wherein non-hexagons are involved in a portion, and the side lengths and internal angles can be also 
quite different from those of regular polygons.
We also observed that the typical diameter of the polygons depends on the locality; particularly,
the typical diameters are about 2 m at O-site and a few centimeters at Sm-site, with a difference by two orders of magnitude. At Y-site and Sz-site, the typical diameters amount to 1 m and 40 cm, respectively (see Table \ref{table_01}).
The total number of polygons acquired by drone shooting at each investigation site was 
1069, 894, 1012, and 3987 at O-site, Y-site, Sm-site, and Sz-site, respectively.
Based on the photos, we calculated the coordinates, side lengths, and internal angles of the polygon vertices using an image analysis software named ArcGIS (Esri).

%----------------
%----------------

% %**********************************************************
\subsection{Quantification of geometric irregularity}
% %**********************************************************

To quantify the geometric deviation of constituent polygons from regular ones, we proposed the following hypothesis.
As the cracks grow, the inner angles of Y-shaped junctions tend to be adjusted so that
all opposing sides of each polygon are evenly spaced;
namely, the typical column diameter is assumed to approach the characteristic crack spacing
determined by the maximum energy release rate.
This hypothesis can be fulfilled by a regular honeycomb lattice,
which is 
the optimal side configuration that enables a significant energy release owing to small fracture energy
needed to create new crack surface along the sides,
provided all polygons are hexagons.
In reality, however, a significant proportion of non-hexagons is involved,
which hinders the ideal evenly spaced side configuration.
Nevertheless, 
the tendency to reduce the local mechanical energy 
(strain energy plus newly created surface energy) regulates the junction angle 
to achieve as large amount of energy release as possible.
Subsequently, the ratio of the polygon area to the total side length enclosing the area
is enlarged.

%----------------
\begin{figure}[ttt]
\centering
\includegraphics[width=0.48\textwidth]{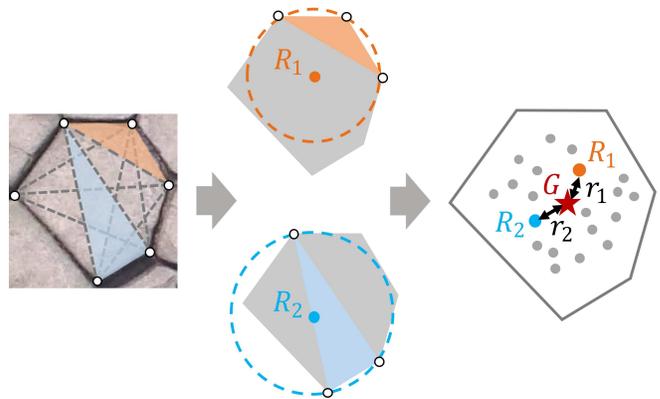}
\caption{Schematic of the calculation method of the degree of geometric deviation of a polygon from a cyclic polygon.
Only two circumcenters $\bm{R}_1$ and $\bm{R}_2$ among $\bm{R}_i$ $(1\le i \le N)$ are depicted.}
\label{fig_03}
\end{figure}
%----------------

Under the abovementioned assumption, 
we quantified the geometric irregularity of constituent polygons through the following procedure (see also Fig.~\ref{fig_03}). 
Assuming $n$-sides with fixed lengths are connected to form an $n$-sided polygon, the corresponding area is maximized for a cyclic polygon ({\it i.e.,} a polygon inscribed in a circle). 
Therefore, the degree of the polygon geometric fluctuation can be quantified by examining the degree of deviation of 
the vertices from a reference circle ({\it i.e.,} the degree of similarity of the polygon to a cyclic polygon).
To this aim, given an $n$-sided polygon obtained from the aerial photo, we selected three vertices
and then evaluated the center of the circumscribed circle that passes through the selected points. 
This calculation was repeated $N$ times (with $N \equiv {n \choose 3}$) for all $N$ triangles comprising the $n$-gon and a set of $N$ circumcenters, designated by $\bm{R}_i$ $(1\le i \le N)$, was obtained.
Then, we evaluated the center of the $N$-point set defined as $\bm{G}= \sum_{i=1}^{N} \bm{R}_i/N$ and
calculated the distances $r_i=|\bm{R}_i-\bm{G}|$ $(1\le i \le N)$ and their average $r_{\rm av} = \sum_{i=1}^N r_i/N$. 
Finally, we defined the degree of deviation of the $n$-gon from the corresponding cyclic polygon, $\chi$, as
\begin{equation}
\chi=\frac{s}{r_{\rm av}}-1,\;\;\; s = \sqrt{\sum_{i=1}^{N}\frac{r_i^2}{N}}.
\label{eq_001}
\end{equation}
Note that $\chi$ 
approximates 
zero when the $n$-gon is almost inscribed in a circle,
while it assumes a large positive value when the $n$-gon deviates considerably from a cyclic polygon (see the Appendix B).

%\color{red}
% %***************************************************************************
\subsection{Voronoi-based simulation of the maturing evolution}
% %***************************************************************************

\begin{figure}[ttt]
\centering
\includegraphics[width=0.45\textwidth]{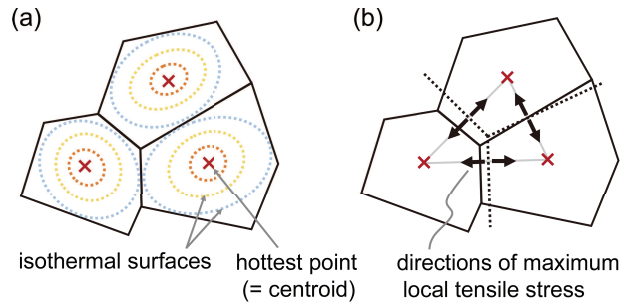}
\caption{
%\color{red}
Diagram of crack evolution process defined in Ref.~\cite{Budkewitsch1994}.
(a) Within the transverse section across adjacent columns, the lava temperature is the highest at the center of each polygon,
marked by the ``hottest point''.
(b) As cooling progresses, a novel crack front is expected to be generated along the dotted lines, i.e., the vertical bisectors of
the neighboring two hottest points, because the local maximum tensile stress occurs in the direction connecting the two hottest points.
%\color{black}
}
\label{fig_4}
\end{figure}

In the geophysical community, Voronoi tessellation has been considered to be particularly useful in modeling 
the crack patterns of actual columnar joints \cite{Budkewitsch1994,Smalley1966}.
The relevance of Voronoi tessellation to real cracking patterns is supported by the following ideas \cite{Budkewitsch1994}:
i) Within a two-dimensional transverse section across the columns at the crack front, the maximum local tensile stress
is expected to occur in the direction parallel to the line segments that connect the centers of adjacent polygons, and
ii) shrinkage cracks are likely to emerge at the midpoints of these line segments, perpendicular to these segments, 
similar to the drawing of vertical bisectors between Voronoi seed points, to relieve the accumulating tensile stress. 
Figure~\ref{fig_4} illustrates the stepwise evolutionary nature of a columnar joint pattern, suggested in Ref.~\cite{Budkewitsch1994},
which demonstrates the rationale for describing this evolution via Voronoi tessellation.

%----------------
%----------------

To compare the real field observation data with Voronoi-based simulation results,
we computed the values of $\chi$ (defined by Eq.~(\ref{eq_001})) of the Voronoi polygons 
and examined the dependence of $\chi$-values on the degree of maturation.
First, we generated Voronoi tessellation using randomly distributed seed points
and matured the tessellation using Lloyd's algorithm \cite{Lloyd1982} through the following steps:

(1) Compute the centroid $z_i (i=1,2,\cdots,n)$ for each

\qquad$n$ Voronoi polygon;

(2) Define a new set of seed points $y_i$ as the centroid

\qquad $z_i$;

(3) Create a new Voronoi tessellation using the new set

\qquad of seeds $y_i$;

(4) Return to Step (1).

\noindent
At each iteration of Steps (1)--(4), we computed the value of $\chi$ of all Voronoi polygons
to examine its dependence on the degree of maturation.
The numerical implementation was performed in MATLAB (MathWorks, Inc.).

%\color{black}

% %**********************************************************
\section{Results}
% %**********************************************************
\subsection{Similarity of probability distributions for $\chi$}
% %**********************************************************

%----------------
\begin{figure}[ttt]
\centering
\includegraphics[width=0.45\textwidth]{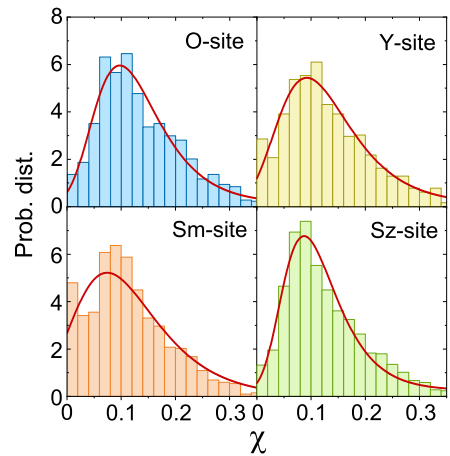}
\caption{Discrete probability distribution of the distortion variable $\chi$ at the four investigation sites.
A fit of each locality data to a Gumbel law given by Eq.~(\ref{eq_005}) was obtained by the nonlinear least-square method.}
\label{fig_05}
\end{figure}
%----------------

\begin{figure}[ttt]
\centering
\includegraphics[width=0.45\textwidth]{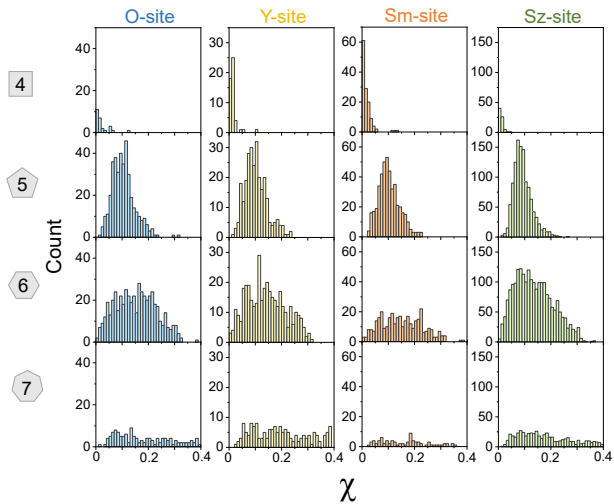}
\caption{Occurrence frequency of $\chi$ for each set of $n$-gons ($n=4,5,6,7$) obtained at the four investigation sites.}
\label{fig_07}
\end{figure}

Figure \ref{fig_05} shows the discrete probability distribution of the $\chi$ value for each of the four investigation sites. 
All the distributions are upward convex and slightly slanted to the left. 
The salient observation is that all four histograms look quite similar,
exhibiting comparable peak heights at almost the same position and wide tails.
Interpreting this similarity is far from trivial. As the polygon sizes 
and lithologic composition of the rocks differ greatly from site to site (see Table \ref{table_01}), 
one may expect that the distribution of $\chi$ will also differ from site to site.
However, our results show that the distribution of $\chi$ is surprisingly similar at all the four sites.
A minor difference in the four distributions is the presence of the left-most isolated peak at $\chi=0$
in the Sm-data (also the Y-data) and the absence of it in the other two; the origin of 
the isolated peak will be clarified in the discussion of Fig.\ref{fig_07}.

\begin{figure}[ttt]
\centering
\includegraphics[width=0.45\textwidth]{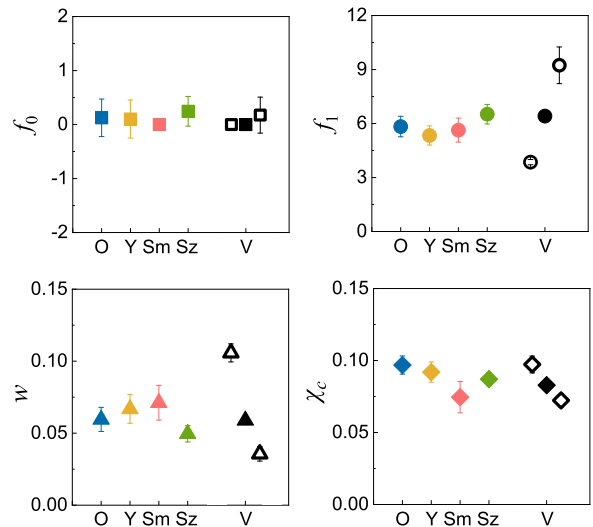}
\caption{
Obtained values for the four fitting parameters. Error bars indicate confidence intervals.
%\color{red}
V represents the results for the Voronoi tessellation at $it_{\rm num}$ = 0, 1, 100 from left to right.
%\color{black}
}
\label{fig_06}
\end{figure}

The common statistical behavior of $\chi$ implies a hidden universality regarding the geometric fluctuation in polygons
appearing at real columnar joints.
In fact, it is demonstrated in Fig.~\ref{fig_05} that all four histograms can be well fitted 
by a special class of extreme value distributions, 
called the Gumbel distribution \cite{GumbelBook1958,Bramwell2009} (solid curves).
The Gumbel distribution is a probability distribution with a density function:
\begin{equation}
f(\chi) = f_0 + f_1 e^{-z} e^{1-e^{(-z)}},
\;\;
z=\frac{\chi-\chi_c}{w},
\label{eq_005}
\end{equation}
where $\chi_c$ indicates the peak position at which $f(\chi)$ has the maximum value of $f_0+f_1$,
$w$ is the peak width, and $f_0$ is the offset such that $f(\chi)\to f_0$ at $\chi\to \infty$.
In the field of fracture mechanics, the Gumbel distribution is known to describe the probability distribution 
of the strength of brittle materials,
as well as other deterioration phenomena and accidental fracture phenomena over time \cite{Anderson2018}.
Figure \ref{fig_06} shows the confidence intervals of the four fitting parameters for each investigation site
%\color{red}
(The Voronoi simulation results, shown in Fig.~\ref{fig_06}, are discussed in section III-D).
%\color{black}
The intervals largely overlap
%\color{red}
for the four investigation sites,
%\color{black}
which supports the robustness of
the functional form of the Gumbel law for the columnar joint.
This is the main result of the present work.

%----------------
%----------------

% %**********************************************************
\subsection{Relevance to generalized Gumbel law}
% %**********************************************************

We also tested a possibility that our data fits an alternative class of Gubmel distribution, 
called a generalized Gumbel distribution (GGD).
The GGD is known to describe non-Gaussian fluctuation phenomena 
in a variety of seemingly unrelated systems,
including turbulent flows \cite{Bramwell1998,Pinton1999,Portelli2003},
viscous fluids \cite{Planet2009}, liquid crystals \cite{Goldburg2001,TothKatona2003,Joubaud2008}
and granular materials \cite{Kou2017}.
It was empirically and analytically \cite{Bramwell2009} suggested that in these fluctuating systems,
the temporal fluctuation of a spatially averaged quantity, $\chi^*$, 
follows a GGD
defined by
\begin{equation}
f_a(\chi^*) = \frac{a^a}{w_a \Gamma(a)} \left\{ e^{- z_a} e^{-e^{(-z_a)}} \right\}^a, \;\; z_a=\frac{\chi^*-\chi^*_a}{w_a}.
\label{eq_024}
\end{equation}
Here,
\begin{equation}
w_a = \frac{\sigma}{\sqrt{\psi'(a)}}\;, \;\;
\chi^*_a = \langle \chi^* \rangle + w_a \left\{ \log a - \psi(a) \right\}.
\label{eq_025}
\end{equation}
and 
$\psi = \Gamma'/\Gamma$, where $\sigma^2$ is the variance of $\chi^*$,
$\langle \chi^* \rangle$ is the average of $\chi^*$,
$\Gamma$ is the Gamma function.
In Eq.~(\ref{eq_024}), $a$ is the only free parameter, and it depends on the system.
Particularly, when $a=1$, the function form of $f_a$ is reduced to that of $f$ given in Eq.~(\ref{eq_005}).

Table \ref{table_01} summarizes the nonlinear square fitting results of our $\chi$ data to Eq.~(\ref{eq_024}).
The obtained values of $a$ are fairly close to $a=1.0$ for the four sites, considering wide margins of error due to
the limited number of polygon samples on the order of $\mathcal{O}(10^3)$.
This result may indicate the validity of GGD-based interpretation of real columnar joint patterns,
while the physical meaning of $a$ remains to be clarified.
%%
%%\color{black}
%%

\begin{table}[bbb]
  \caption{Optimal value of parameter $a$ for the GGD fitting.}
  \centering
  \begin{tabular}{lccc}
    \hline
    Site  & $a$  & Lithology & Mean diameter \\
    \hline \hline
    O-site  & 0.85 $\pm$ 0.52 \hspace*{3pt} & Andesite     & 1.78 m \\
    Y-site  & 0.88 $\pm$ 0.51 \hspace*{3pt} & Trachybasalt & 94.4 cm \\
    Sm-site & 1.54 $\pm$ 1.78 \hspace*{3pt} & Rhyolite     & 5.22 cm \\
    Sz-site & 0.57 $\pm$ 0.29 \hspace*{3pt} & Andesite     & 42.2 cm \\
    \hline
  \end{tabular}
\label{table_01}
\end{table}

%----------------
%----------------

% %*******************************************************************
\subsection{Contribution to $\chi$ from different $n$-gon sets}
% %*******************************************************************

The isolatted peak at $\chi=0$ shown in Fig.~\ref{fig_05} for Sm-site 
as well as the wide error margin of the $a$ value (see Table \ref{table_01})
can be explained
by considering the contribution
from each set of $n$-gons for different $n$ to the $\chi$ distribution, as plotted in Fig.~\ref{fig_07}.
Note that the vertical axis indicates the frequency of appearance, instead of the probability,
in order to show the difference in the total number of $n$-gons observed for different $n$.
The histogram for $n=4$ exhibits a sharp peak close to $\chi=0$ for all investigation sites.
This possibly occurs because fluctuation of the quadrilateral shape is severely restricted. 
In fact, when the distance between opposite sides happens to be small,
the energy release rate is significantly reduced \cite{Jenkins2009}. 
Therefore, as maturation proceeds, the distance between opposite sides in all quadrilaterals is expected to approach 
the characteristic length determined by effective energy release; subsequently, the quadrilateral shape will approximate a cyclic quadrilateral.
Specifically, at Sm-site, linearly long cracks and T-shaped junctions, which may have been created during the initial fracture stage,
persist in the outcrop despite the long maturing process, as clearly observed in Fig.~\ref{fig_01}.
These T-shaped junctions may promote the generation of quadrilaterals,
which results in the pronounced peak at $\chi=0$ observed in the Sm-data shown in Fig.~\ref{fig_05}.
In contrast, the value of $\chi$ for $n=7$ is broadly scattered with an almost constant distribution.
This is because when $n$ is large, 
the internal angle at a vertex can be occasionally close to $180^\circ$ or a bit larger, 
causing a quasi-T-shaped branch (see Appendix C).
Because the two sides extending from this quasi-T-shaped branch should line up almost linearly,
the circumcenter of the triangle containing this branch point deviates significantly 
from the circumcenter of the other triangles.
As a consequence, $\chi$ for $n$-gons with $n\ge 7$ exhibits significant fluctuation.

%\color{red}

% %**********************************************************
\subsection{Comparison of Voronoi simulation model and real columnar joints}
% %**********************************************************

\begin{figure}[ttt]
\centering
\includegraphics[width=0.45\textwidth]{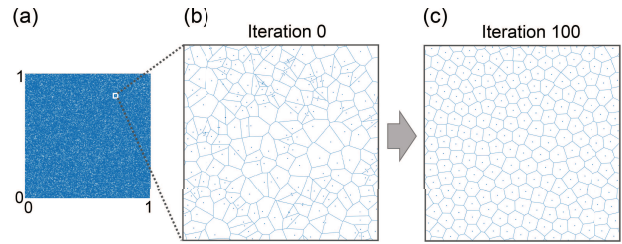}
\caption{
%\color{red}
Simulation results of Voronoi tessellation: (a)(b) initial condition, (c) 100th iteration.
%\color{black}
}
\label{fig_08}
\end{figure}

To generate the Voronoi tessellation, we used the open-source program
developed by Burkardt \cite{Burkardt2015}.
For this MATLAB code, it is necessary to set the number of iterative steps ($it_{\rm num}$),
number of generators ($g_{\rm num}$), and number of sample points 
in the unit square considered to estimate the Voronoi regions ($s_{\rm num}$).
In the actual simulations, we varied $it_{\rm num}$ from 0 to 100 by setting
$g_{\rm num}$ = $1.0\times10^{5}$ and $s_{\rm num}$ = $1.0\times10^{7}$.
Figures \ref{fig_08} (a) and (b) show the result of Voronoi tessellation at $it_{\rm num}$ = 0;
after 100 iterations ($it_{\rm num}$ = 100), we obtained a more regular Voronoi tessellation,
as shown in Fig.~\ref{fig_08} (c).

%----------------
\begin{figure}[ttt]
\centering
\includegraphics[width=0.45\textwidth]{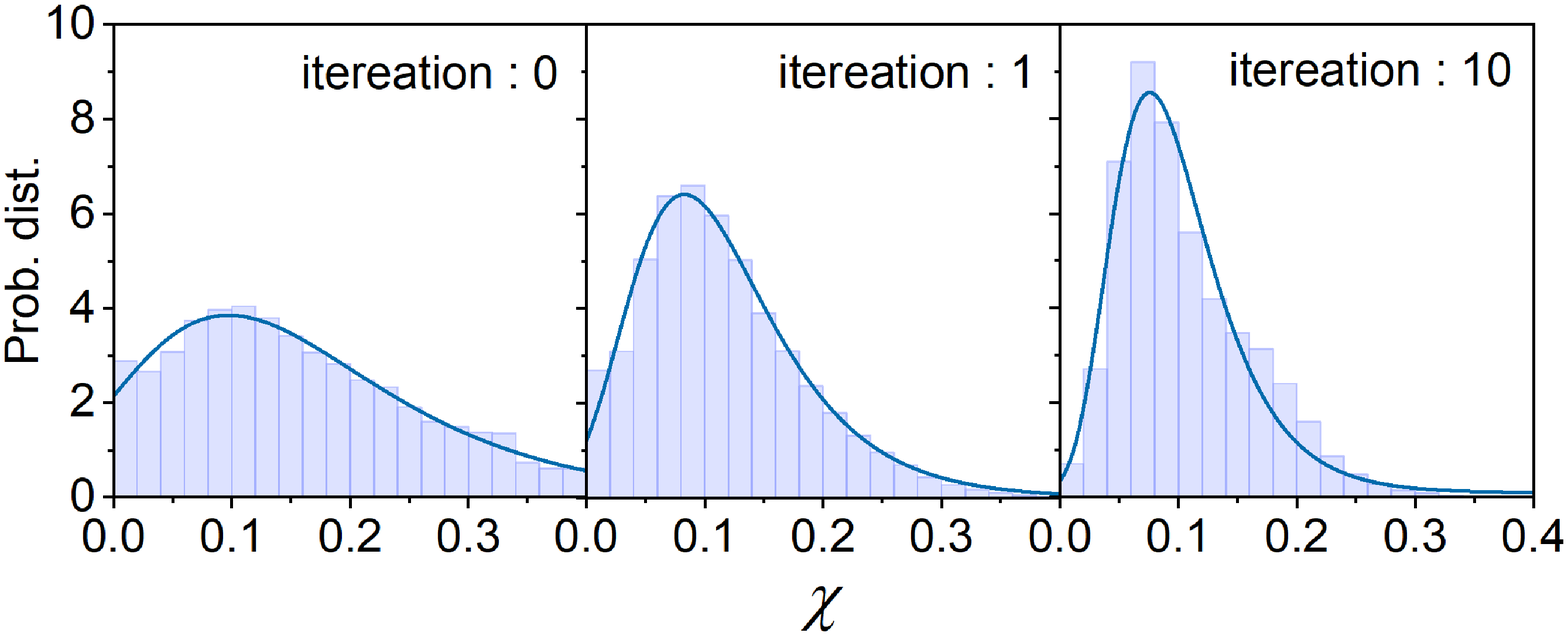}
\caption{
%\color{red}
Discrete probability distribution of the distortion variable $\chi$ at the Voronoi tessellation. The number of iterative steps is 0, 1, and 10.
The fit of each iteration data point to the Gumbel law specified by Eq.~(\ref{eq_005}) is obtained using the nonlinear least-squares method.
%\color{black}
}
\label{fig_09}
\end{figure}

%----------------
%%
The results of the discrete probability distribution of the $\chi$ value for the Voronoi tessellation
created for each iteration are shown in Fig.~\ref{fig_09}.
As in the case of the field observation data, 
all the distributions are upward convex and slightly inclined to the left.
In particular, for the histogram with one iteration ($it_{\rm num}$ = 1),
both the peak height and tail width are 
in quantitative agreement with those of the field measurement data.
Moreover, the three histograms obtained by the Voronoi simulations can be well fitted 
by the Gumbel distribution, as shown by the curves in Fig. \ref{fig_09}.

\begin{figure}[ttt]
\centering
\includegraphics[width=0.45\textwidth]{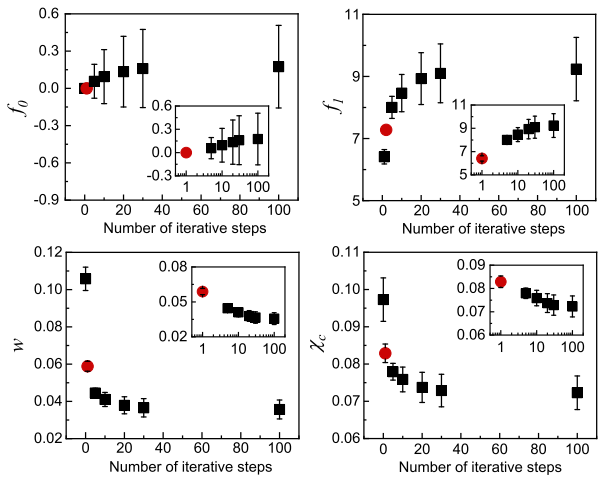}
\caption{
%\color{red}
Values of the four fitting parameters for each iterative step.
Only the data points at one iteration are highlighted by circles (marked in red).
Error bars indicate confidence intervals.
Inset: Semi-logarithmic plot showing the nearly exponential behavior of the data.
%\color{black}
}
\label{fig_10}
\end{figure}

Figure \ref{fig_10} shows the four fitting parameters of the Voronoi tessellation 
for various iterations ranging from 0 to 100.
All the parameters change monotonically as the iteration proceeds,
in a nearly exponential manner, and eventually converge to constant values.
To compare the simulation results with the field observation data,
we selected the four fitting parameters at $it_{\rm num}$ = 0, 1, and 100, as shown in Fig.\ref{fig_10},  and
plotted them in Fig. \ref{fig_06}; the values correspond to the leftmost three data points in each panel of Fig.\ref{fig_06}.
Clearly, the fitting parameter values of the Voronoi tessellation at
$it_{\rm num}$ = 1(solid symbols) are most similar to those of the actual columnar joints.
The same conclusion is drawn if the data points at $it_{\rm num}$ = 1 are
replaced by those at $it_{\rm num}$ = 2, 3 or slightly larger values(not shown).
The results shown in Figs. \ref{fig_06} and \ref{fig_09} validate
the theoretical hypothesis that the real columnar joints matured according to the Voronoi process,
and the degree of maturation is limited to that attained by one or a few iterative numerical simulations.
%%
%\color{black}

%----------------
%----------------

%----------------
%----------------

%% %**********************************************************
\section{Discussion}
%% %**********************************************************

It is expected that the Gumbel law applies not only to the solidified lava flow
but also to fracturing systems that show polygonal cracking in general \cite{Goehring2013,SBTang2017}.
For instance, columnar joint-like prismatic patterns can be reproduced by table-top analogue experiments 
using starch-water mixtures \cite{Akiba2017,XMa2019,Goehring2006,Goehring2009,Mizuguchi2005, Nishimoto2007}.
Other candidates are 
two-dimensional polygonal fractures observed at the surface of 
mud and clay after repeated drying cycles \cite{Goehring2010}
and polygonal terrain that has undergone an annual thermal cycle \cite{Domokos2020},
both of which are dominated by Y-shaped junctions.
Considering the similarity in the cracking mechanism, 
it is possible that the Gumbel distribution governs 
the cross-sectional shape of this polygonal cracking;
partial experimental verifications on the issue will be presented in future work.

%\color{red}
The numerical simulations quantitatively demonstrated that 
the statistical properties of the distortion variable $\chi$ for real columnar joints are
consistent with those deduced from the Voronoi-based maturing process.
This result supports the existing hypothesis
that the formation process of columnar joints can be explained by the coupling of 
the isothermal distribution and crack propagation \cite{Budkewitsch1994}.
Moreover, we could reproduce the geometric fluctuation of the columnar joints with the Voronoi tessellation
in only a few iterations.
A large number of 
iterations is not required for this reproduction owing to the following reason:
In the case of lava, fractures propagate slowly, and
thus, a geological timescale is required to complete
the maturation process; this timescale corresponds to only one
iteration in the context of Voronoi simulations.
Moreover,
in reality, the maturation is likely inhibited by external environmental factors
(e.g., changes in cooling conditions owing to diastrophism and sea level changes).
The fusion of a few adjacent columns into one \cite{DeGraff1993},
which was not considered in the numerical simulations,
may also affect the evolution of the $\chi$-distribution.
%\color{black}

Before closing the article, we emphasize that
the existence of a common rule for the statistics of polygon geometry over different geological sites is totally nontrivial. 
The lava we investigated has different chemical compositions and different crack length scales. 
In addition, environmental conditions (erosion of solidified lava and heat dissipation, for instance) 
as well as physical conditions (spatial distribution of stress field and temperature field, for instance) during the columnar joint formation 
should be also different for each investigation sites. 
Therefore, there is no trivial reason for our measurement data from different sites to follow a single particular class of probability distribution. 
The present study also revealed that a series of environmental and physical factors listed above, 
which are locality-dependent, can be excluded as a dominant factor for realizing the universal Gumbel law. 
Neither lithographic composition nor crack length scale should be the dominant factor as they are also different between investigation sites. 
These exclusions may provide a guideline for identifying the dominant factor of the universal fluctuation law 
using numerical simulations or analogue experiments.

%%**********************************************************
\section{Conclusion}
%%**********************************************************

In conclusion, we have demonstrated that polygonal fractures observed at the outcrop of several columnar joints are commonly described by the Gumbel distribution function.
The robustness of the Gumbel law, governing the spatial fluctuation of the vertex configuration in each polygonal cross-section, 
is a surprising finding considering the diversity of the field conditions in terms of the column size, locality, and lithologic composition. 
We expect that this discovery will help establish a new classification of polygonal crack patterns that are ubiquitous in nature.

%%**********************************************************
\section*{Acknowledgment}
%%**********************************************************

This work was financially supported by JSPS KAKENHI Grants 
(grant numbers 18H03818, 19K03766, and 20J10344).

\appendix
%%*************************************************************
\section{Area of a polygon inscribed in a circle}
%%*************************************************************

In Appendix A, we prove the following theorem:
{\it the area of a convex polygon with sides $\ell_i$ ($i=1,\cdots,n$) is maximized when the polygon is inscribed in a circle. 
The maximum area is independent of the order of the sides $\{\ell_i\}$.} 
The following proof is based on the argument presented by Hewes in Ref.~\cite{Hewes1948}.
For the sake of simplicity, we first show that the above theorem holds for convex quadrilaterals and then extend it for $n$-sided convex polygons with $n \ge 5$.

%----------------
\begin{figure}[ttt]
\centering
\includegraphics[width=0.25\textwidth]{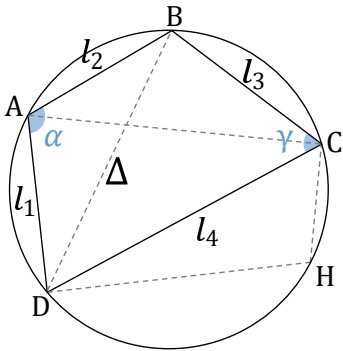}
\caption{A convex quadrilateral ABCD
and a convex pentagon ABCHD formed by adding a point H to the quadrilateral.}
\label{fig_11}
\end{figure}
%----------------

Suppose the quadrilateral with sides $\ell_i$ ($i=1,2,3,4$) depicted in Fig.~\ref{fig_11}.
We assume that the length of each side $\ell_i$ is fixed.
The order in which the sides are connected is arbitrary, 
and the angle between the sides can be arbitrarily chosen as long as the resulting quadrangle is convex.
The area $S$ of the quadrilateral can be computed from
\begin{equation}
S = \frac12 \left( \ell_1 \ell_2 \sin \alpha + \ell_3 \ell_4 \sin \gamma \right). \tag{A1}
\label{eq_008}
\end{equation}
Here, $\alpha$ and $\gamma$ are the angles between sides $\ell_1$ and $\ell_2$ and sides $\ell_3$ and $\ell_4$, respectively.
Using the two angles, the length of the diagonal line $\Delta$ can be expressed as follows:
\begin{equation}
\Delta = \left( \ell_1^2 + \ell_2^2 - 2 \ell_1 \ell_2 \cos \alpha \right)^{\frac12}
= \left( \ell_3^2 + \ell_3^2 - 2 \ell_3 \ell_4 \cos \gamma \right)^{\frac12}. \tag{A2}
\label{eq_009}
\end{equation}
Using the identity $\sin^2 \gamma + \cos^2 \gamma = 1$, we can eliminate $\gamma$
from Eqs.~(\ref{eq_008}) and (\ref{eq_009}).
We then obtain
\begin{equation}
\left( \frac{2S - p \sin \alpha}{q} \right)^2 + \left( \frac{k-2p \cos\alpha}{2q} \right)^2 = 1. \tag{A3}
\label{eq_010}
\end{equation}
Here, $p=\ell_1 \ell_2$, $q=\ell_3 \ell_4$, and $k=\ell_1^2+\ell_2^2-\ell_3^2-\ell_4^2$ are constants. 
Therefore, to maximize the quadrilateral area, we must have
\begin{equation}
\frac{dS}{d\alpha} = \frac{p}{4} \left(\frac{k \sin\alpha - 4S \cos\alpha}{p \sin\alpha - 2S} \right) = 0, \tag{A4}
\end{equation}
which implies $S = (k \tan\alpha)/4$.
Substituting it to Eq.~(\ref{eq_010}), we obtain
\begin{equation}
\cos\alpha = \frac{k}{2(p+q)} = \frac{\ell_1^2+\ell_2^2-\ell_3^2-\ell_4^2}{2 \left( \ell_1 \ell_2 + \ell_3 \ell_4 \right)}. \tag{A5}
\label{eq_012}
\end{equation}
From Eqs.~(\ref{eq_009}) and  (\ref{eq_012}),
we deduce that $\cos \gamma = -\cos \alpha$, {\it i.e.,} $\alpha + \gamma = \pi$.
This corresponds to the necessary and sufficient condition for the given quadrilateral with sides $\ell_i$ ($i=1,2,3,4$) 
to be inscribed in a circle.

We next consider a convex pentagon with sides $\ell_i$ ($i=1,2,3,4,5$), as depicted by ABCHD in Fig.~\ref{fig_11}. 
If the pentagon area is maximized, then quadrilateral ACHD area is also maximized,
because the area of triangle ABC is uniquely determined.
It thus follows that quadrilateral ACHD is inscribed in a circle.
To determine whether the circumscribing circle passes by point B,
we consider that when the area of the unchanged pentagon is maximized, then the area of quadrilateral BCHD is also maximized.
Therefore, quadrilateral BCHD should be inscribed in the circle defined by points D, H, and C.
Finally, we conclude that, if the pentagon area is maximized, the pentagon is inscribed in a unique circle.
Similarly, the inscribed property can be extended to an $n$-gon with $n\ge 6$.

%*************************************************************
\section{The limiting value of $\chi$ for a cyclic polygon}
%%*************************************************************

In Appendix B, we show that $\chi$ defined by Eq.~(\ref{eq_001})
converges to zero
in the limit at which the polygon approaches infinitely a cyclic polygon.
Without loss of generality, we restrict our argument to the case of a quadrilateral with a certain symmetry;
the argument can be extended straightforwardly to other general $n$-gons with $n\ge 4$.

%----------------
\begin{figure}[ttt]
\centering
\includegraphics[width=0.38\textwidth]{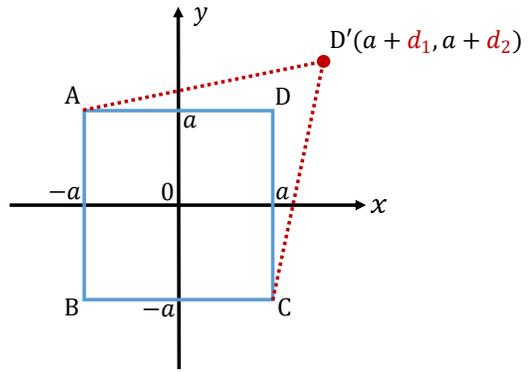}
\caption{A square ABCD centered at the origin with a side length of $2a$.}
\label{fig_12}
\end{figure}
%----------------

First, we consider a square ABCD, as shown in Fig.~\ref{fig_12}, centered at the origin with a side length of $2a$.
Next, the position of vertex D is displaced by $d_1$ and $d_2$ in the $x$ and $y$ directions, respectively, and labeled D'.
The coordinate of the circumcenter of triangle ABD', written as $\bm{R}_1 = (x_1, y_1)$,
can be expressed as
\begin{equation}
x_1 = \frac{K a}{2} \left( \frac{1}{2+\delta_1} \right), \quad y_1 = 0, \tag{B1}
\end{equation}
where 
\begin{equation}
K = \delta_1^2 + 2\delta_1 + \delta_2^2 +2\delta_2,\;\;
\delta_1 = \frac{d_1}{a}, \;\;
\delta_2 = \frac{d_2}{a}. \tag{B2}
\end{equation}

Similarly, the coordinates of the circumcenters of triangles BCD', ACD', and ABC
given by $\bm{R}_2$, $\bm{R}_3$, and $\bm{R}_4$, respectively, are expressed as
\begin{align}
x_2 &= 0, \quad y_2 = \frac{K a}{2} \left( \frac{1}{2+\delta_2} \right), \tag{B3}\\
x_3 &= y_3 = \frac{K a}{2} \left( \frac{1}{2+\delta_1+\delta_2} \right), \tag{B4}\\
x_4 &= y_4 =0. \tag{B5}
\label{eq_021}
\end{align}
Using these results, the center of the four circumcenters 
defined by $\bm{G} = \sum_{i=1}^4 \bm{R}_i/4 = (x_G, y_G)$ becomes
\begin{align}
x_G &= \frac{Ka}{8} \left( \frac{1}{2+\delta_1} + \frac{1}{2+\delta_1+\delta_2} \right), \tag{B6}\\
y_G &= \frac{Ka}{8} \left( \frac{1}{2+\delta_2} + \frac{1}{2+\delta_1+\delta_2} \right). \tag{B7}
\end{align}
It is readily proven that every $r_i=|\bm{R}_i-\bm{G}|$ $(i=1,2,3,4)$ has the form of
\begin{equation}
r_i = \frac{K a}{8} \eta_i. \tag{B8}
\end{equation}
Here, $\eta_i$ is a function of $\delta_1$ and $\delta_2$,
whose functional forms depend on $i$.
As a consequence, $r_{\rm av} = \sum_{i=1}^4 r_i/4$ and $s = \sqrt{ \sum_{i=1}^{4} r_i^2/4}$
are given by
\begin{equation}
r_{\rm av} = \frac{K a}{8} \left( \frac{\sum_{i=1}^4 \eta_i}{4} \right), \quad 
s = \frac{K a}{8} \sqrt{ \frac{\sum_{i=1}^4 \eta_i^2}{4} }, \tag{B9}
\end{equation}
respectively, and thus
\begin{equation}
\chi=\frac{s}{r_{\rm av}}-1 \;\;=\;\; 
\frac{2 \sqrt{ \sum_{i=1}^4 \eta_i^2 }}{\sum_{i=1}^4 \eta_i} - 1. \tag{B10}
\label{eq_app_05}
\end{equation}

Assuming that the displacement of point $D$ is infinitesimally small so that $\delta_1 \ll 1$ and $\delta_2 \ll 1$,
we expand $\sqrt{ \sum_{i=1}^4 \eta_i^2 }$ and $\sum_{i=1}^4 \eta_i$
in terms of $\delta_1$ and $\delta_2$,
and keep the terms up to the first order.
Then, Eq.~(\ref{eq_app_05}) can be approximated by
\begin{equation}
\chi= \frac{2 (8 -3 \delta_1 -3 \delta_2)}{ 16  -6 \delta_1 -6 \delta_2} - 1 = 0. \tag{B11}
\label{eq_app_06}
\end{equation}

This result completes the proof.

%*************************************************************
\section{Remark on quasi-T-shaped branches}
%%*************************************************************

%----------------
\begin{figure}[ttt]
\centering
\includegraphics[width=0.2\textwidth]{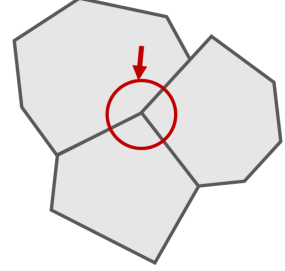}
\caption{Schematics of a quasi-T-shaped branches.}
\label{fig_13}
\end{figure}
%----------------

%----------------
\begin{figure}[ttt]
\centering
\includegraphics[width=0.32\textwidth]{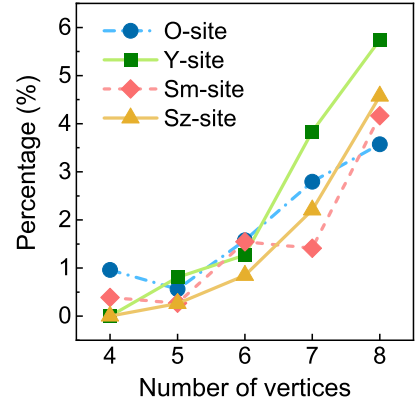}
\caption{Occurrence frequency of $160^\circ$-$170^\circ$ and $190^\circ$-$200^\circ$ internal angle
for each $n$-gons ($n=4,5,6,7,8$) obtained at the four investigation sites.}
\label{fig_14}
\end{figure}
%----------------

In Appendix C, we provide an auxiliary explanation of quasi-T-shaped branches. As mentioned in the main text, 
the internal angle at a vertex of $n$-gons for large $n$ can sometimes takes a value close to $180^\circ$, resulting 
in a quasi-T-shaped branch as marked in the schematics of Fig.~\ref{fig_13}. 
Such the quasi-T-shaped branches are difficult to be distinguished from real T-shaped branches composed of a linear long crack.
Therefore, in our image analysis, Y-branches containing the internal angles whose values are in the range of $170^\circ$ to $190^\circ$ 
were identified as T-branches, and these branch points were not included in vertices of the corresponding $n$-gons. 
According to this criterion, for example, the upper left polygon depicted in Fig.~\ref{fig_13} is regarded as a hexagon, not a heptagon.

Figure \ref{fig_14} shows the occurrence frequency of large internal angles within each set of $n$-gons at different investigation sites. 
The vertical axis indicates the percentage of the number of internal angles whose values are in the range of $160^\circ$ to $200^\circ$ 
(with those close to $180^\circ$ being excluded under the abovementioned criterion).
It follows that the large internal angles are more likely to appear in $n$-gons with larger $n$, which causes the broad distribution of 
$\chi$ for polygons with $n\ge 7$ as presented in Fig.~\ref{fig_07}.

%%%%%%%%%%%%%%%%%%%%%%%%%%%%%%%%%%%%%%%%%%
\bibliographystyle{apsrev4-1}
\bibliography{AkibaUniversal2021}
%%%%%%%%%%%%%%%%%%%%%%%%%%%%%%%%%%%%%%%%%%]

\end{document}